\documentclass[11pt]{article}
\usepackage{amsmath}
\usepackage{amssymb}
\usepackage{graphicx}
\usepackage{multicol}
\usepackage{indentfirst}
\usepackage{caption}
\usepackage{placeins}

\usepackage{calc}
\usepackage[margin=0.5in]{geometry}

\usepackage{color}
\usepackage[usenames]{xcolor}

\usepackage[cp1252]{inputenc}
\usepackage{listings}
\definecolor{m_green}{RGB}{63,127,95}
\definecolor{m_blue}{RGB}{42,0,255}
\definecolor{m_purple}{RGB}{127,0,85}
\title{Enumeration of Moir\'e Patterns of a Hexagonal Twisted Bilayer and Intercalated Transition Metals in Twisted h-BN}
\author{Matthew Ciesler, Damien West, Shengbai Zhang}
\date{%
    \textit{Department of Physics, Rensselaer Polytechnic Institute, NY}\\%
    \today
}
\begin{document}
\maketitle

\begin{abstract}\noindent
A real-space method using generating integers is used to classify the possible moir\'e patterns for two equal hexagonal lattices. The result is that the rotations that take $(n,m)$ to $(m,n)$ with $n,m$ relatively prime form the fundamental moir\'e transformations, and the number of lattice coincidence areas within each supercell is given by $(n-m)^2$. The scheme may be extended to cases where the lattice constants differ. Additionally, we consider a system with a transition metal between the layers of a twisted bilayer of h-BN. We find that the lowest energy configurations for such an arrangement are those at aligned and anti-aligned sites of the moir\'e pattern, depending on the transition metal, and the low-symmetry sites possess high magnetization. 
\end{abstract}

\begin{multicols}{2}
\section{Introduction}

Monolayer physics has attracted considerable interest in recent studies. Two dimensional materials offer some unique properties---for example, the band structure of graphene contains Dirac points and cones \cite{malko2012competition} \cite{neto2009electronic}. Beyond an isolated monolayer, systems of interest include a monolayer adsorbed onto a substrate and multiple monolayers stacked together \cite{liu2016van} \cite{novoselov20162d}. The substrate-monolayer case is especially important because of the efforts towards growing graphene on different substrates \cite{tetlow2014growth}. The combination of different monolayers is also promising because different monolayers can be engineered together analogously to Lego bricks, forming a Van der Waals heretostructure \cite{geim2013van}. These emerging research areas demonstrate why it is important to have a complete, fundamental understanding of the geometry of these systems.

When two lattices are brought together, as in the above examples, a superlattice structure called a moir\'e pattern emerges, with varying amounts of periodicity \cite{yankowitz2019tuning}\cite{tarnopolsky2019origin}\cite{zhang2020flat}. These twisted bilayers have demonstrated exotic properties such as superconductivity \cite{oh2021evidence}\cite{po2018origin}\cite{an2020interaction}. In cases considered here, we will assume that the moir\'e pattern does have some cell for which the pattern is truly periodic. Note that it has been demonstrated that in real systems, the lattices may be incommensurate and thus do not have such a repeating symmetry \cite{ortix2012graphene}. 

The moir\'e pattern will depend on the twist angle, and the Bravais lattice types and lattice parameters of each of the layers. On top of this, each layer can respond with a certain amount of strain, for example to accommodate for slightly incommensurate lattice constants \cite{artaud2016universal}. These complicating factors will be addressed later, but for now let us consider the simple case of two identical hexagonal lattices being brought together.

\section{Methods}

\subsection{Real-space Construction of Irreducible Moir\'e Patterns}
For what follows, only the details of the lattice will be important, so the results can be applied equally to systems such as hexagonal BN or graphene. As shown in Fig. \ref{fig:hexlattice}, we choose the basis $\mathbf{a}_1=a(1,0)$ and $\mathbf{a}_2=a\left(\frac12,\frac{\sqrt3}2\right)$. (Basis angles of $60^\circ$ \cite{tkatchenko2007commensurate} and $120^\circ$ \cite{zeller2014possible} are equivalent.) A point in this lattice can be described with coordinates $\mathbf{n}=(n_1,n_2)$, such that its position is $\mathbf{r}_\mathbf{n}=n_1\mathbf{a}_1+n_2\mathbf{a}_2$.

\begin{center}
\includegraphics[width=.25\textwidth]{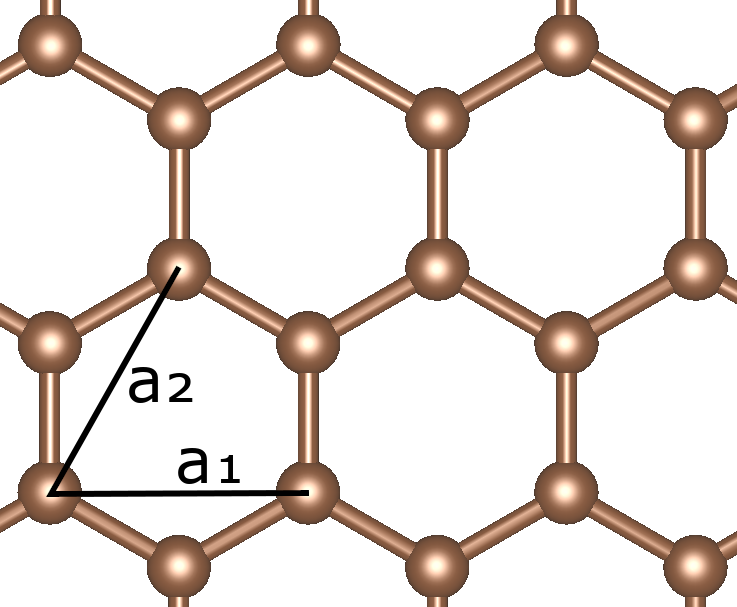}
\end{center}
\captionof{figure}{Basis for the hexaonal lattice used for formulating the Moir\'e Condition Equation.}
\label{fig:hexlattice}
\bigskip

In order to generate a twisted bilayer moir\'e pattern, we must first choose some center of rotation. Suppose the center of rotation is $\mathbf{r}_0$, and the angle of rotation is $\theta$. Then the positions of the rotated cell are
$$\mathbf{r}'=\mathbf{r}_0+R_\theta\left(\mathbf{r}-\mathbf{r}_0\right)$$
$$\mathbf{r}'=R_\theta\mathbf{r}+\left(R_\theta-I\right)\mathbf{r}_0.$$

Notice that the effect of choosing a different center of rotation only shifts the other layer. Since this shift does not affect the periodicity of the moir\'e pattern, we can choose an arbitrary center of rotation. Therefore, we select $\mathbf{r}_0$ to be at a lattice site. With this choice of basis, a nontrivial moir\'e pattern will exist whenever there is an atom in the lattice which is rotated to a symmetrically inequivalent site on the same lattice. Clearly this is only possible if the distances to each of the sites are the same.

Suppose that the first site has coordinates $\mathbf{n}$ and the second site has coordinates $\mathbf{m}$. Then the condition can be written
$$\left|n_1\mathbf{a}_1+n_2\mathbf{a}_2\right|^2=\left|m_1\mathbf{a}_1+m_2\mathbf{a}_2\right|^2.$$
Note that
$$\left|n_1\mathbf{a}_1+n_2\mathbf{a}_2\right|^2\neq n_1^2\left|\mathbf{a}_1\right|^2+n_2^2\left|\mathbf{a}_2\right|^2,$$
because the basis vectors are not orthogonal. Expanding the original equation and simplifying, we obtain the Moir\'e Condition Equation (MCE)
$$n_1^2+n_1n_2+n_2^2=m_1^2+m_1m_2+m_2^2.$$
Approaching this equation directly corresponds to working in real space, contrasted with working in reciprocal space \cite{zeller2014possible}.

\subsection{Reduction of Moir\'e Transformations}

The MCE is a Diophantine Equation, which is an equation whose solutions are integers; its solutions form what is known as the hexagonal number sequence \cite{tkatchenko2007commensurate}. It is similar in spirit to the Pythagorean equation $a^2+b^2=c^2$, whose integer solutions are the well-known Pythagorean triples, e.g. $3^2+4^2=5^2$. Since the MCE is a homogeneous quadratic Diophantine Equation, its solutions can be enumerated by means of generating integers, in a very similar fashion to that of the the Pythagorean equation. Using the method of parameterization, one finds that integers $(k,t_1,t_2,t_3)$ enumerate the solutions according to
\begin{align*}
n_1^* &= -t_1^2+t_1t_3+t_2^2-t_3^2 \\
n_2^* &= -2t_1t_2-t_2^2+t_2t_3 \\
m_1^* &= -2t_1t_3-t_2t_3+t_3^2 \\
m_2^* &= t_1^2+t_1t_2+t_2^2-t_3^2 \\
\left(\mathbf{n},\mathbf{m}\right) &= \frac{k\cdot\left(\mathbf{n}^*,\mathbf{m}^*\right)}{\mathrm{gcd}(n_1^*,n_2^*,m_1^*,m_2^*)}.
\end{align*}
However, the solutions $(\mathbf{n},\mathbf{m})$ that are generated from the generating integers $(k,\mathbf{t})$ are not in 1:1 correspondence. There is degeneracy---many different combinations of the generating integers will result in the same solution $(\mathbf{n},\mathbf{m})$. Also, solutions of the MCE $(\mathbf{n},\mathbf{m})$ which share the same twist angle are geometrically equivalent.

The twist angle $\theta$ is given by 
$$\mathbf{r}_\mathbf{n}\cdot\mathbf{r}_\mathbf{m}=\left|\mathbf{r}_\mathbf{n}\right|\:\left|\mathbf{r}_\mathbf{m}\right|\:\cos(\theta).$$
Substituting in terms of $\mathbf{n}$ and $\mathbf{m}$, this gives
$$\cos\theta=\frac{2n_1m_1+2n_2m_2+n_1m_2+n_2m_1}{2(n_1^2+n_1n_2+n_2^2)}.$$
In terms of the generating integers, this is
$$\cos\theta=\frac{3t_1(t_3-t_2)}{2(t_1^2-t_1(t_3-t_2)+(t_3-t_2)^2)}-\frac12.$$
Note that the twist angle has lost a degree of freedom, since it only depends on $t_1$ and $t_3-t_2$. Additionally, since the angle is independent of $k$, and values of $k$ greater than unity only cause a scale increase in the area of the unit cell, we may fix $k=1$. This has the advantage of minimizing the cell volume, which is desired for computations.

There are still more degeneracies to be addressed. Since the hexagonal lattice is symmetric under $120^\circ$ rotations, it suffices to consider $\theta$ modulo $120^\circ$. That is, we restrict $\theta$ to be in the range $-60^\circ\leq\theta\leq60^\circ$. Another symmetry in this bilayer corresponds to permuting the two layers, or flipping the whole system like a pancake. This symmetry takes $\theta\mapsto-\theta.$ Therefore, the angle range can be shrunk down to $0^\circ\leq\theta\leq60^\circ$.

Still, we may have two sets of integers $(\mathbf{n},\mathbf{m})$ and $(\mathbf{n}',\mathbf{m}')$ which give the same moir\'e twist angle but are not symmetrically equivalent under the above operations. As an explicit example (see Fig. \ref{fig:moirepattern}), a twist angle of $\theta=\arccos\left(\frac{73}{74}\right)\approx9.43^\circ$ is obtained for all of the following MCE solutions $(\mathbf{n},\mathbf{m})$:
\begin{align*}
&(4, 3, 3, 4) \\
&(15, 2, 13, 5) \\
&(19, 5, 16, 9) \\
&(26, 1, 23, 6).
\end{align*}

Since these transformations share the same angle, they correspond to the same physical moir\'e pattern. Therefore, there must be some relationship that demonstrates that these $(\mathbf{n},\mathbf{m})$ are equivalent.

In fact, we can show that the higher-order $(\mathbf{n},\mathbf{m})$ transformations are a supercell of the first one, which we will call the primitive transformation (for this particular angle). To motivate the relationship that we will show, first let us try to express the second vector $(15, 2, 13, 5)$ in terms of a coordinate system based on the primitive transformation $(4, 3, 3, 4)$.

The basis for the coordinate system will have an angle of $60^\circ$, just like the underlying coordinate system. Therefore, for the pre-transformed system, the basis is $\{(4, 3),(-3, 7)\}$, where the second vector is formed by rotating the first counter-clockwise by $60^\circ$. These coordinates are with respect to the $\{\mathbf{a}_1,\mathbf{a}_2\}$ basis. In this basis, we can express
$$\begin{bmatrix}15\\2\end{bmatrix}=3\begin{bmatrix}4\\3\end{bmatrix}-\begin{bmatrix}-3\\7\end{bmatrix}=\begin{bmatrix}4 & -3\\3 & 7\end{bmatrix}\begin{bmatrix}3\\-1\end{bmatrix}.$$
In the transformed system, we can repeat this process, yielding
$$\begin{bmatrix}13\\5\end{bmatrix}=3\begin{bmatrix}3\\4\end{bmatrix}-\begin{bmatrix}-4\\7\end{bmatrix}=\begin{bmatrix}3 & -3\\4 & 7\end{bmatrix}\begin{bmatrix}3\\-1\end{bmatrix}.$$
Note that, in general, we would not expect these coordinates to be integer, let alone the same integer values. The fact that this has occurred is not a coincidence---it is because the $(15, 2, 13, 5)$ transformation is a $3\times-1$ supercell of the primitive transformation $(4, 3, 3, 4)$.

One may ask---it always the case that higher-order transformations can be reduced to a primitive transformation, and, if so, what form does the primitive transformation take? We claim that the answer is yes, and the primitive transformation may take the form $(n, m, m, n)$, where $n$ and $m$ are relatively prime. To show this, we search for integers $(h_1,h_2)$ and $n,m$ which define the supercell for some transformation $(\mathbf{n},\mathbf{m})$. They must satisfy
$$\begin{bmatrix}n_1 \\ n_2\end{bmatrix}=\begin{bmatrix}n & -m \\ m & n+m\end{bmatrix}\begin{bmatrix}h_1\\h_2\end{bmatrix}$$
$$\begin{bmatrix}m_1 \\ m_2\end{bmatrix}=\begin{bmatrix}m & -n \\ n & n+m\end{bmatrix}\begin{bmatrix}h_1\\h_2\end{bmatrix}$$
In terms of $g=\mathrm{gcd}(n_1^*,n_2^*,m_1^*,m_2^*)$ and generating integers, this is
$$\frac1g\begin{bmatrix}-t_1^2+t_1t_3+t_2^2-t_3^2\\-2t_1t_2-t_2^2+t_2t_3\\-2t_1t_3-t_2t_3+t_3^2\\t_1^2+t_1t_2+t_2^2-t_3^2\end{bmatrix}=\begin{bmatrix}n & -m \\ m & n+m\\m & -n \\ n & n+m\end{bmatrix}\begin{bmatrix}h_1\\h_2\end{bmatrix}$$
which simplifies to
$$\frac1{3g}\!\!\begin{bmatrix}
(t_1 + 2 t_2 - 2 t_3) (t_1 + 2 t_2 + t_3)\\
-(2 t_1 + t_2 - t_3) (t_1 + 2 t_2 + t_3)\\
-(t_1 + 2 t_2 - 2 t_3) (2 t_1 + t_2 - t_3)\\
(2 t_1 + t_2 - t_3)^2
\end{bmatrix}\!\!=\!\!\begin{bmatrix}
n & 0\\
m & 0\\
0 & n\\
0 & m
\end{bmatrix}\!\begin{bmatrix}h_1\\h_2\end{bmatrix}$$

This shows that, as long as $(2 t_1 + t_2 - t_3)/g\equiv 0 \mod 3$, then an integral solution exists with $n,m,h_1,h_2$ identified from the equation above. In this procedure, we only considered the primitive transformation $(n,m,m,n)$, when in fact there are other symmetrically equivalent transformations that would give a similar result. These other cases exhaust the other possibilities in the condition above.

Since this transformation holds, the solution $(\mathbf{n},\mathbf{m})$ is either symmetrically equivalent to a primitive solution or a supercell of such a solution. Since we always want the smallest unit cell for doing analysis and computations, we can restrict out attention to just the primitive solutions. That is to say, each essentially different moir\'e pattern is equivalent to the twist given by $(n_1,n_2,m_1,m_2)=(n,m,m,n)$. Furthermore, since each combination $(n_1,n_2,m_1,m_2)=(n,m,m,n)$ solves the MCE, they categorize the essentially different moir\'e patterns. However, note that the solutions $(n,m,m,n)$ that have $n-m\equiv0\mod3$ can be reduced to a $(n',m',n'+m',-m')$ subclass with a smaller unit cell. The $(n,m,m,n)$ and $(n,m,n+m,-m)$ Moir\'e patterns represent a reflection across the $60^\circ$ line and $0^\circ$ line, respectively.

\subsection{Geometrical interpretation of $(n,m)$}

As shown in Fig. \ref{fig:higherorder}, there is a geometric meaning behind the difference $n-m$. If $n-m=1$, then the twist angle will be small, and each moir\'e supercell will have one lattice coincidence area, also known as a moiron \cite{hermann2012periodic}. When the difference is increased, the twist angle increases and each supercell no longer has a single moiron in it. In fact, one can see from Fig. \ref{fig:higherorder} that each unit cell has $(n-m)^2$ moirons in it. Also, as $n$ and $m$ are increased, the volume of the moir\'e supercell increases. These observations are consistent with other treatments of Moir\'e patterns \cite{zeller2017indexing}.

\begin{center}
\includegraphics[width=.325\textwidth]{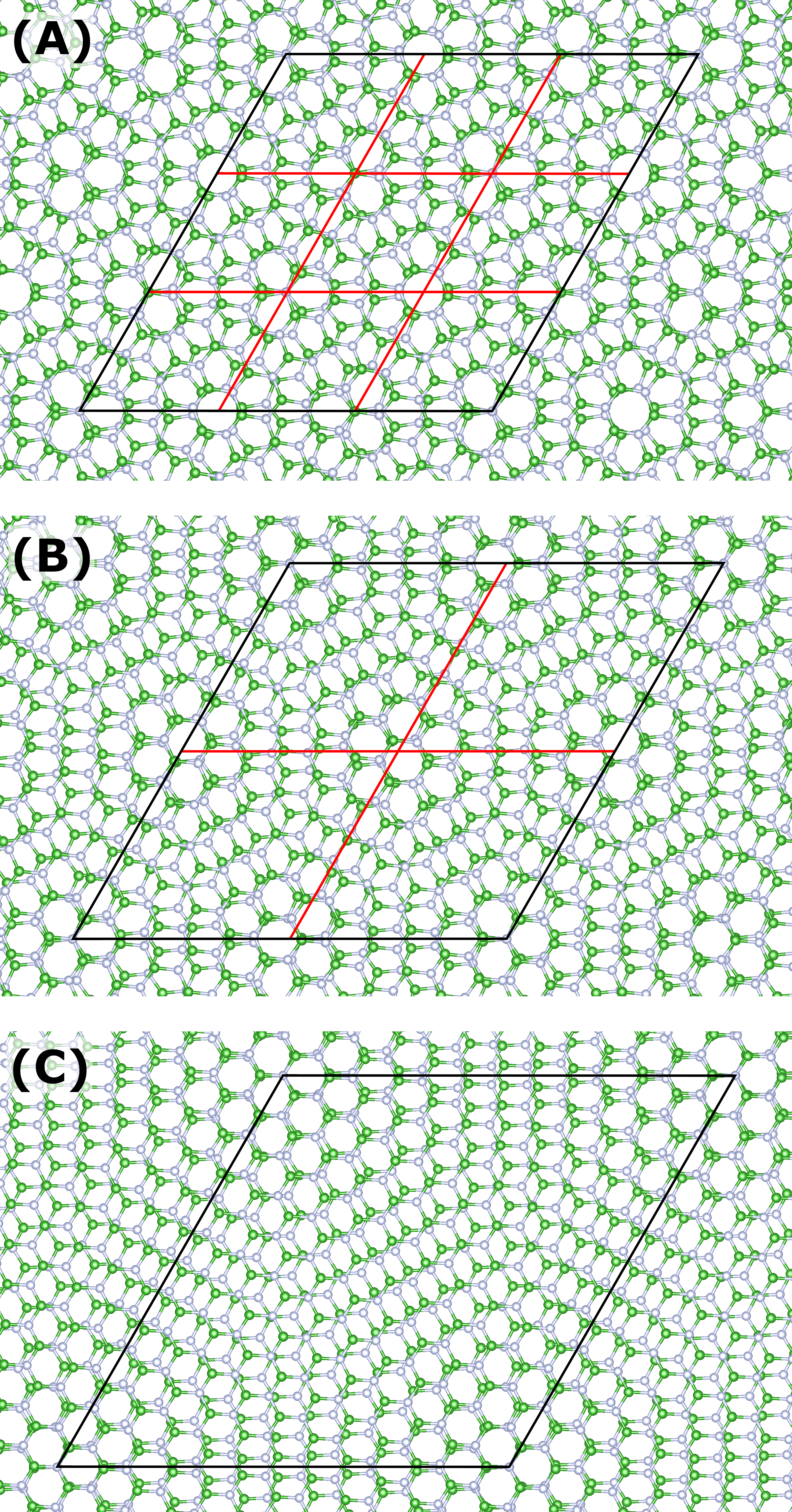}
\end{center}
\captionof{figure}{Moir\'e pattern formed by the solutions, from top to bottom: (A) $(7,4,4,7)$, (B) $(7,5,5,7)$, and (C) $(7,6,6,7)$. The black cell corresponds to the supercell, and the red subsections represent symmetrically similar subregions. In the case of (B), the lattice coincidence areas (moirons \cite{hermann2012periodic}) of different subregions are not equivalent to each other, but overall do repeat periodically. For (A) there is partial degeneracy of different regions, such that a smaller unit cell can be constructued for this Moir\'e pattern (see text).}
\label{fig:higherorder}
\bigskip

\begin{center}
\includegraphics[width=.325\textwidth]{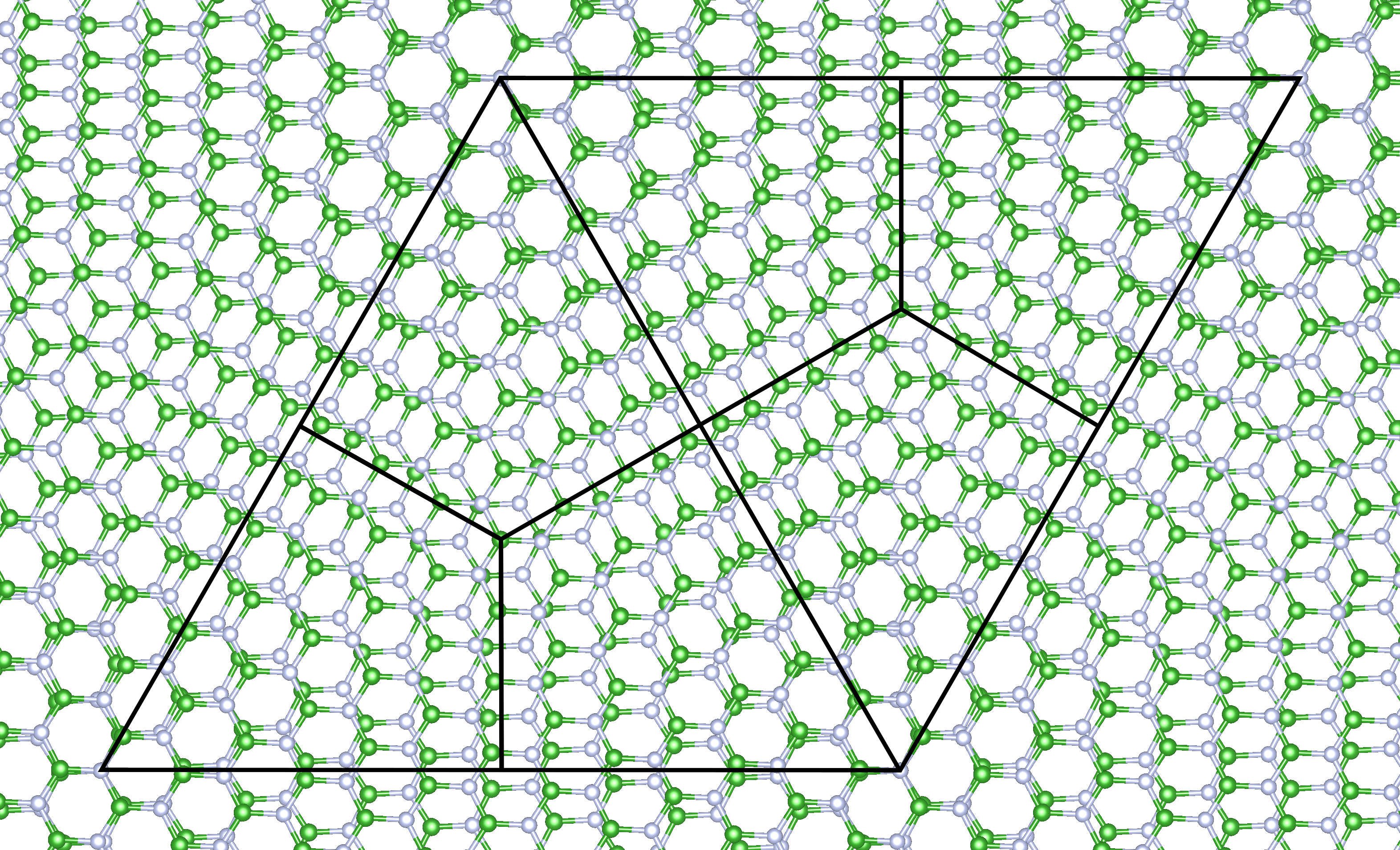}
\end{center}
\captionof{figure}{Symmetrical subregions of the Moir\'e pattern. If the two atoms are the same, each region will be symmetrically equivalent to all the other regions. Otherwise, one region and its reflection (e.g. bottom left and top right) will be inequivalent, but will together be equivalent to the other four regions.}
\label{fig:symmetry_reduction}
\bigskip

\subsubsection{Non-Identical Lattice Parameters}

This method could be accommodated for two lattices which have different lattice parameters. If $a_1\neq a_2$, then the distance criterion implies

$$a_1^2\left(n_1^2+n_1n_2+n_2^2\right)=a_2^2\left(m_1^2+m_1m_2+m_2^2\right).$$

If $a_1$ and $a_2$ are taken to be real-valued quantities, this expression will in general have no integer solutions. However, if we let $p/q$ be a rational approximation to $a_1^2/a_2^2$, then we may replace the real numbers with their rational counterparts and obtain
$$p\left(n_1^2+n_1n_2+n_2^2\right)=q\left(m_1^2+m_1m_2+m_2^2\right),$$
which is a Diophantine equation just like before. The $p$ and $q$ are fixed constants, and a similar procedure can be used to generate $(\mathbf{n},\mathbf{m})$. The approximation of the real lattice constants by these rational numbers manifests as small amounts of strain in order to commensurate the lattices. The rational numbers can be chosen from truncating the continued fraction of $a_1^2/a_2^2$. A balance must be struck regarding these integers---if $p,q$ are too large, then the moir\'e supercell is unwieldy for computations, but if they are too small, then the strain values are unphysically large.


\subsection{First-principles calculations}

The energetics, electronic properties, and magnetic moments of period 4 transition metals (TMs), intercalated within a h-BN twisted homobilayer were determine through first-principles calculations based on density functional theory within the local density approximation. These calculations were performed in periodic supercells containing a vacuum spacing of 10.4 \AA and a bilayer twist angle of approximately $9.43^\circ$ corresponding to $(n,m)=(4,3)$ and containing 74 B and 74 N atoms. Interactions between ion cores and valence electrons are described by the projector augmented-wave (PAW) method \cite{PhysRevB.50.17953, PhysRevB.59.1758}, as implemented in the VASP package \cite{PhysRevB.54.11169, PhysRevB.59.1758}. Plane waves with a kinetic energy cutoff of 400 eV were used as the basis set. As a lateral shift of the bilayers relative to each other can change the local chemical environment of a hexagonal site, care needed to be taken for structural relaxation of the TM intercalated bilyaer system. To investigate the energetics associated with the TM trapped at local minima in the moir\'e pattern, several bilayer atoms (most distant from the TM site) where constrained in the lateral direction in order to prevent relaxation of the system to the global energy minimum. Non-collinear spin was used for all calculations and constrained structural relaxation was performed until the maximum Hellman-Feynmann force on any atom in an unconstrained degree of freedom was below 0.01 eV/\AA. 

\section{Results}

To investigate how the different local symmetry within the Moir\'e pattern associated with a homobilayer can affect preferential bonding and the electronic characteristics of defects, we investigated the series of period 4 transition metals (TMs) at different sites intercalated in twisted bilayer h-BN. Intercalated impurities, such as the ones studied here, have been prepared experimentally \cite{ichinokura2016superconducting}\cite{kanetani2012intercalated}\cite{sugawara2011fabrication}, and have potential applications such as superconductivity \cite{huang2015prediction}, quantum computing \cite{sokolov20202d}, and energy storage \cite{ji2019lithium}. The $(n,m)=(4,3)$ h-BN supercell, shown in Fig. \ref{fig:moirepattern}, was chosen because it is large enough to contain different chemical environments for the transition metal while being small enough to facilitate fast computations.

In order to describe different locations for the transition metal within the twisted bilayer, we shall introduce the concept of honeycomb sites. The purpose of these sites is to enumerate different places in the twisted bilayer that would give the transition metal a different chemical environment. In order to sample these different environments, we introduce a set of sites distributed thoughout the Moir\'e pattern. Each site is located within the plane between the two layers, and above the center of a hexagon of the bottom layer (see Fig. \ref{fig:calcspots}). Note that we have used symmetry to reduce the total number of sampling points.

These points, considered as part of the twisted bilayer, are not in symmetrically special positions. However, each site has a varying degree of local symmetry, and can be categorized into three groups: (a) aligned regions, where a BN hexagon from the top layer and bottom layer roughly coincide with each other; (b) unaligned regions, where the hexagons from the different layers are shifted in a way that makes the relative positions of boron and nitrogen completely asymmetric; and (c) anti-aligned regions, where the hexagons are maximally displaced from each other, with an atom of the top layer nearly directly above the center of the hexagon in the lower layer. The sites closest to the origin are low-numbered and correspond with group (a), medium-numbered with group (b), and high-numbered with group (c).

Each initial atomic state was formed by placing a transition metal at a site, within a pre-relaxed twisted bilayer. Upon relaxation, there was a minor amount of bond distortion and buckling of the h-BN. The most significant changes occured when a B or N atom was located directly above or below the transition metal, in which case it was pushed away from the layer by around 0.2 \AA, which is about an order of magnitude larger than the distortions caused by the twisted bilayer.

\begin{center}
\includegraphics{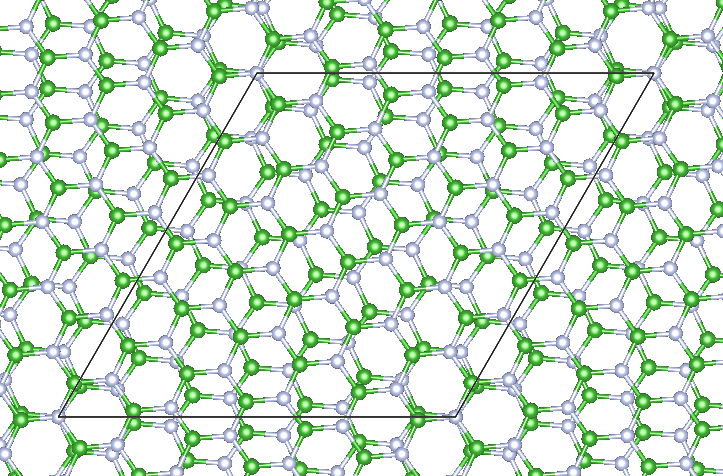}
\end{center}
\captionof{figure}{Moir\'e pattern formed by rotating $(4,3)$ to $(3,4)$, which corresponds to a solution $(\mathbf{n},\mathbf{m})=(4,3,3,4).$ The twist angle for this pattern is $\theta=\arccos\left(\frac{73}{74}\right)\approx9.43^\circ$. The parallelogram shows the unit cell of $(3,4)$ moir\'e supercell.}
\label{fig:moirepattern}
\bigskip

\begin{center}
\includegraphics[width=.4\textwidth]{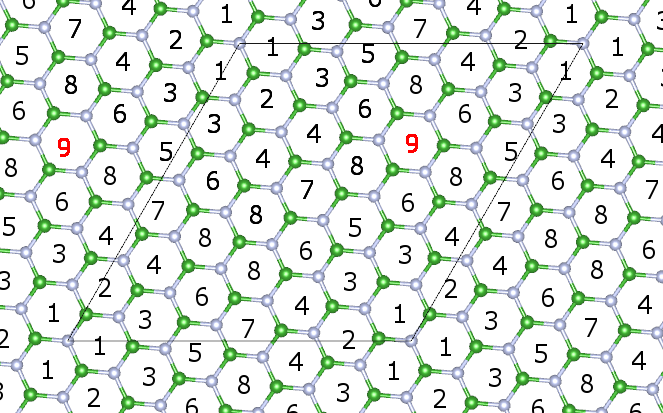}
\end{center}
\captionof{figure}{Transition metal sites, numbered from closest to furthest from a lattice point. The transverse position of the intercalated transition metal is at the center of the hexagons of this layer. The cells with the name number are equidistant from the lattice origin. These sites are used to probe the variation in properties of intercalated transition metals.}
\label{fig:calcspots}
\bigskip

The energy and magnetization results from the sampled transition metals are shown in Table \ref{table:1} and \ref{table:2}, respectively. We first note that zinc has a positive binding energy at all sites; in other words, it is energetically unstable at any intercalated site. For every other atom, the binding energy is negative, but not as low as its respective cohesive energy. For reference, the cohesive energies of these elements (in eV) are approximately given as follows: Sc, -3.90; Ti, -4.85; V, -5.31; Cr, -4.10 eV; Mn, -2.92; Fe, -4.28; Co, -4.39; Ni, -4.44 eV; Cu, -3.49; Zn, -1.35. We can see that for some elements, such as chromium and copper, the most favorable sites are about 2.5 eV higher than bulk, where the difference is less than 1.0 eV in iron, cobalt, and nickel. In most elements, the lowest energy occurs on site 1, where the layers are aligned with each other. Many elements also have a marked increase or decrease in energy at site 9, which is the anti-aligned site. Nickel and copper have minimum energy configurations at an intermediate site, which is unusual because those sites are located where the layers are disorganized. The binding energies at different sites for titanium and manganese are plotted in Fig. \ref{fig:resultbind}.

\end{multicols}
\begin{table*}[bh!]
\centering
\begin{tabular}{|c c c c c c c c c c c|} 
 \hline
 \multicolumn{11}{|c|}{Intercalated Transition Metal Binding Energy in a (3,4) Twisted Bilayer of h-BN} \\
 \hline
 site no. & Sc & Ti & V & Cr & Mn & Fe & Co & Ni & Cu & Zn \\
 \hline
1 & -2.52 & -3.32 & -3.15 & -1.73 & -1.77 & -3.33 & -3.85 & -3.75 & -0.69 & 1.52 \\
2 & -2.43 & -3.22 & -3.02 & -1.60 & -1.68 & -3.26 & -3.83 & -3.79 & -1.05 & 1.53 \\
3 & -2.40 & -3.13 & -2.91 & -1.48 & -1.60 & -3.20 & -3.83 & -3.90 & -1.03 & 1.54 \\
4 & -2.18 & -2.98 & -2.50 & -1.16 & -1.46 & -2.81 & -3.79 & -3.93 & -0.98 & 1.56 \\
5 & -2.17 & -3.09 & -2.59 & -1.30 & -1.43 & -2.71 & -3.68 & -4.14 & -0.89 & 1.56 \\
6 & -2.07 & -2.89 & -2.38 & -1.03 & -1.35 & -2.72 & -3.76 & -4.08 & -0.92 & 1.56 \\
7 & -2.12 & -3.05 & -2.56 & -1.28 & -1.30 & -2.97 & -3.66 & -4.28 & -0.79 & 1.60 \\
8 & -1.74 & -2.71 & -2.19 & -1.05 & -1.56 & -3.12 & -3.71 & -4.09 & -0.70 & 1.55 \\
9 & -2.26 & -3.28 & -2.80 & -1.58 & -1.28 & -2.95 & -3.33 & -4.07 & -0.66 & 1.55 \\
 \hline
\end{tabular}
\caption{Binding energies (in eV) of the intercalated transition metal (ITM) in each site. The values are computed by taking the energy of the relaxed combined ITM system and subtracting the energies of twisted bilayer and the lone transition metal.}
\label{table:1}
\end{table*}

\begin{table*}[h]
\centering
\begin{tabular}{|c c c c c c c c c c c|} 
 \hline
 \multicolumn{11}{|c|}{Intercalated Transition Metal Magnetization in a (3,4) Twisted Bilayer of h-BN} \\
 \hline
 site no. & Sc & Ti & V & Cr & Mn & Fe & Co & Ni & Cu & Zn \\
 \hline
1 & 1.00 & 1.98 & 1.00 & 0.00 & 0.99 & 2.00 & 1.00 & 0.00 & 1.00 & 0.00 \\
2 & 1.00 & 1.99 & 1.00 & 0.00 & 0.99 & 1.99 & 1.00 & 0.00 & 1.00 & 0.00 \\
3 & 1.00 & 2.00 & 1.00 & 0.00 & 0.99 & 2.00 & 1.00 & 0.00 & 0.99 & 0.00 \\
4 & 1.00 & 2.00 & 3.00 & 1.99 & 0.99 & 0.00 & 1.00 & 0.00 & 0.99 & 0.00 \\
5 & 0.99 & 2.00 & 1.00 & 0.00 & 0.99 & 2.00 & 1.00 & 0.00 & 1.00 & 0.00 \\
6 & 0.99 & 2.00 & 1.00 & 0.00 & 0.99 & 2.00 & 1.00 & 0.00 & 1.00 & 0.00 \\
7 & 0.99 & 2.00 & 1.00 & 0.00 & 0.99 & 2.00 & 1.00 & 0.00 & 0.99 & 0.00 \\
8 & 1.00 & 2.00 & 3.00 & 4.00 & 2.99 & 2.00 & 1.00 & 0.00 & 1.00 & 0.00 \\
9 & 1.00 & 2.00 & 1.00 & 0.00 & 0.99 & 2.00 & 1.00 & 0.00 & 0.99 & 0.00 \\
 \hline
\end{tabular}
\caption{Magnetization (in $\mu_B$) of the intercalated transition metal (ITM) in each site.}
\label{table:2}
\end{table*}

\clearpage
\begin{multicols}{2}

The magnetization results also show some patterns but some unusual exceptions as well. The magnetization for one set of calculations for iron and vanadium is plotted in Fig. \ref{fig:resultmag}. As expected, we see that if there are an odd number of $d$-electrons, then there is an odd magnetization, and likewise with an even number of electrons (magnetization). The exceptions to this rule, e.g. Ti at site 1, exist because states in each spin channel are nearly degenerate. On recalculating the magnetization of each site with the exact same structure, we found that multiple magnetization states are possible in some cases.

Site 8, which has low symmetry, has high magnetization states for all the elements except nickel and zinc. At site 9, the transition metal has tetrahedral coordination to boron atoms, and octahedral coordination to nitrogen atoms. At site 8, the geometry is similar to site 9, but with the atom species reversed and more noticeable distortion. At site 1, the transition metal has trigonal prism coordination to both boron and nitrogen, with the prisms rotated from each other by $60^\circ$.

\begin{center}
\noindent\begin{picture}(215,115)
\put(0,0){\includegraphics[width=.35\textwidth]{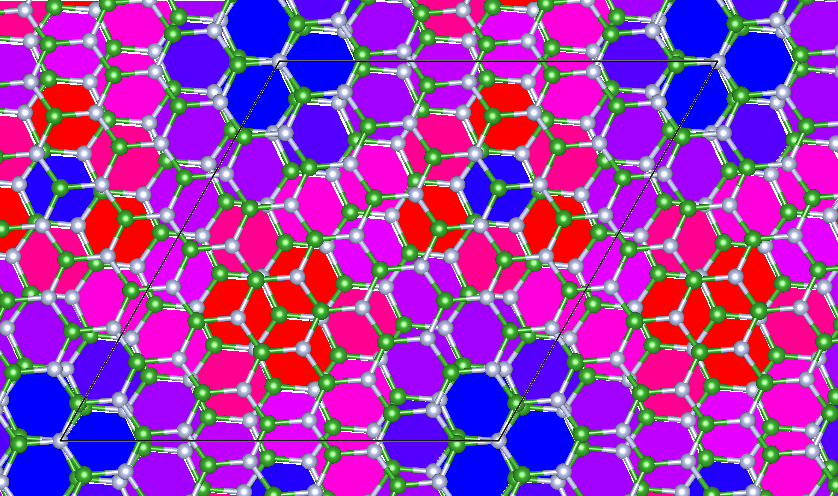}}
\put(195,0){\includegraphics[width=0.016289\textwidth]{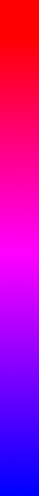}}
\put(207,100){-2.72 eV}
\put(207,5){-3.32 eV}
\end{picture}\vspace{10pt}
\noindent\begin{picture}(215,115)
\put(0,0){\includegraphics[width=.35\textwidth]{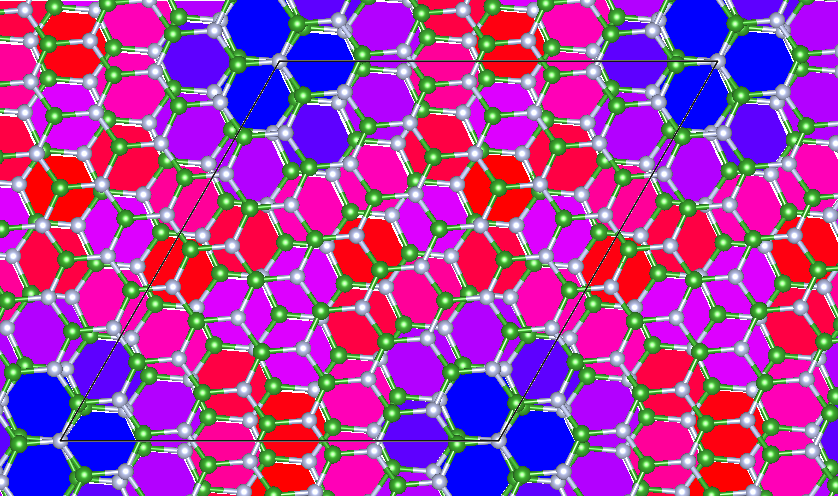}}
\put(195,0){\includegraphics[width=0.016289\textwidth]{plots/img_scale1.png}}
\put(207,100){-1.29 eV}
\put(207,5){-1.78 eV}
\end{picture}
\captionof{figure}{Binding energies for each of the sites shown spatially in a unit cell. The top plot corresponds to titanium, while the bottom plot corresponds to manganese.}
\label{fig:resultbind}
\end{center}

\section{Discussion}
In order to gain some understanding of the dependence of the energy and magnetic moment of the defect with respect to the different sites in the moire pattern, we consider a simple perturbative model. In this model, the degenerate d-orbitals associated with the TM are split by the crystal field associated with their interaction with nearby B or N in their local environment. Here we assume that the perturbing potentials are isotropic, yielding matrix elements of Hamiltonian $H_{ij}=\left\langle\phi_i|\Sigma_k V(r-r_k)|\phi_j\right\rangle$, where k is summed over neighboring atoms, and the potential due to B and N have opposite sign. Assuming a $1/r$ potential for the interaction, the d-orbital splitting calculated from this model at each site is shown in Fig.\ref{fig:hBN-model}.

\begin{center}
\noindent\begin{picture}(215,115)
\put(0,0){\includegraphics[width=.35\textwidth]{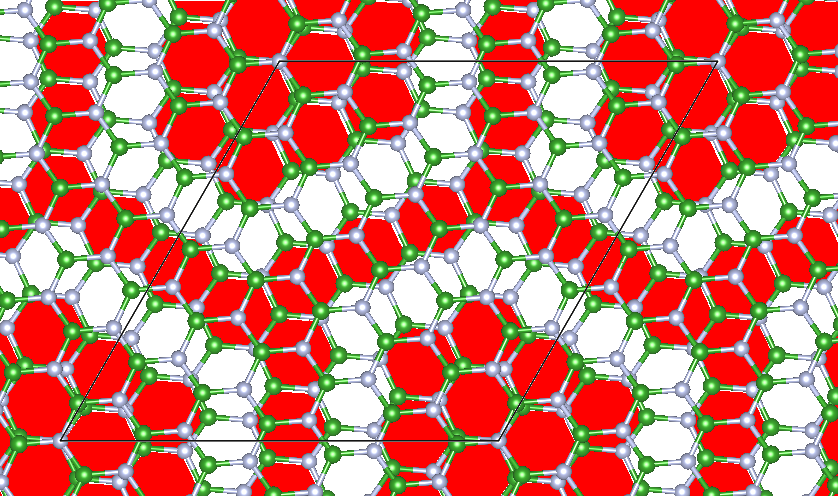}}
\put(195,0){\includegraphics[width=0.016289\textwidth]{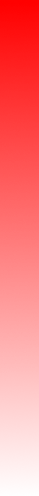}}
\put(207,100){2.0 $\mu_B$}
\put(207,5){0.0 $\mu_B$}
\end{picture}\vspace{10pt}
\noindent\begin{picture}(215,115)
\put(0,0){\includegraphics[width=.35\textwidth]{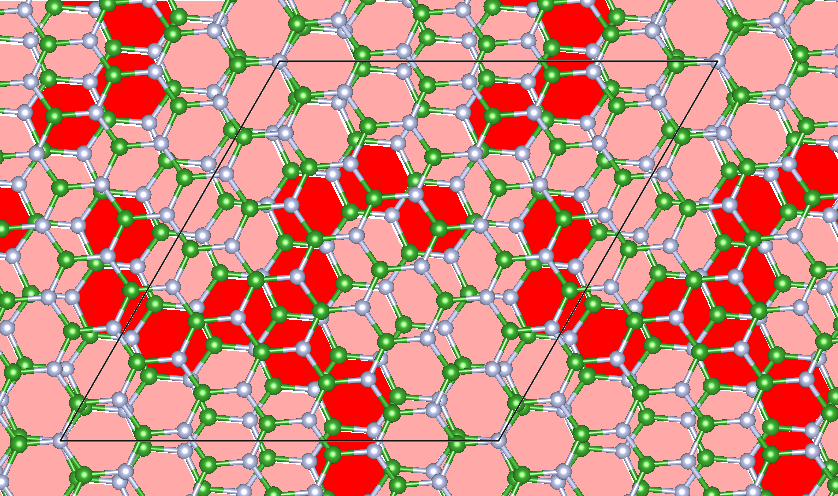}}
\put(195,0){\includegraphics[width=0.016289\textwidth]{plots/img_scale2.png}}
\put(207,100){3.0 $\mu_B$}
\put(207,5){0.0 $\mu_B$}
\end{picture}
\captionof{figure}{Magnetization for each of the sites shown spatially in a unit cell. The top plot corresponds to iron, while the bottom plot corresponds to vanadium.}
\label{fig:resultmag}
\end{center}

\begin{center}
\includegraphics[width=.45\textwidth]{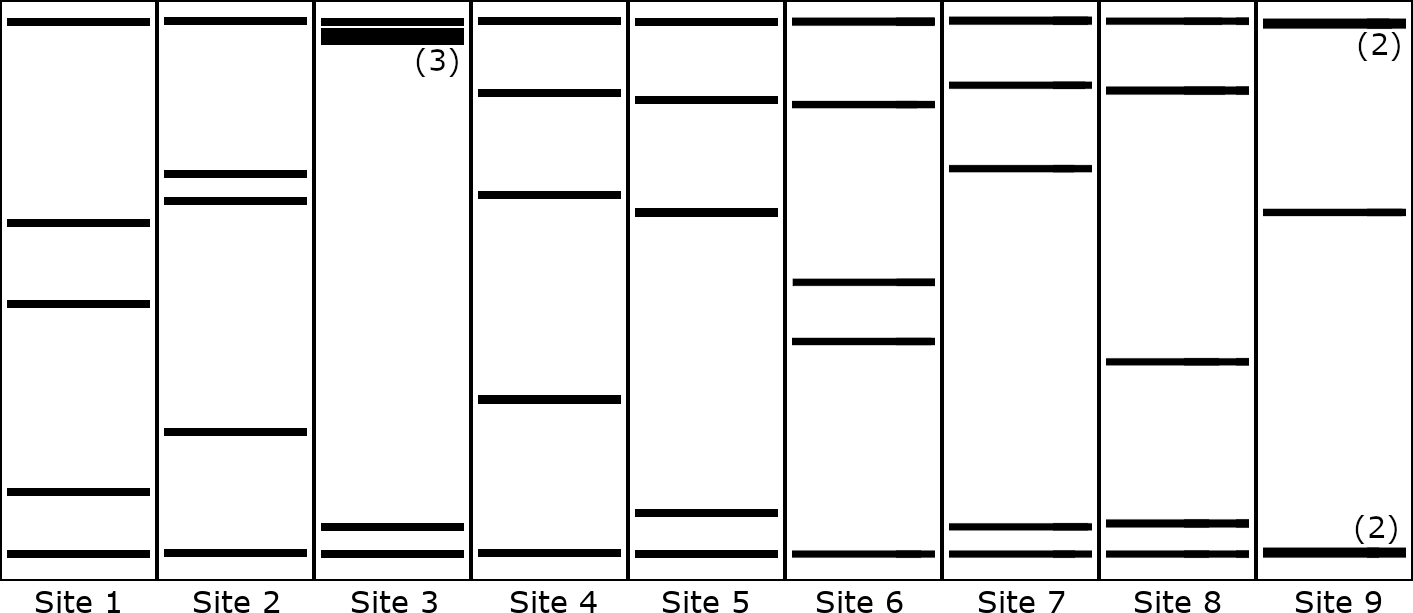}
\end{center}
\captionof{figure}{Results from the crystal field model. For each site, the five eigenenergies are given on a normalized scale. The values are split from degeneracy by the presence of nearby B and N atoms. For clarity, the nearly degenerate groups of states are labeled with a number which indicates the number of states.}
\label{fig:hBN-model}
\bigskip

Despite its simplicity, this model captures much of the physics associated with the different spin states observed in Table 2. Taking the trigonal prismatic geometry associated with site 1, the perturbative model predicts two sets of split doubles with a higher energy singlet, consistent with the result from DFT. With spin-splitting small compared to the crystal-field splitting, one would expect that increasing the number of electrons filling the d-orbital would yield total magnetic moments of 1, 2, 1, 0, 1, 2, 1, 0, 1, and 0 $\mu_B$, respectively. Note, that this is complete agreement with row 1 of Table 2. Furthermore, sites in the immediate vicintiy of site 1, i.e. sites 2 and 3, are only slightly distorted and have similar structure with resulting magnetic moments, leading to a cluster of sites with a similar spin magnetic moments in the vicinity of site 1, as can be seen in either Fig 5. or Fig. 6.

As one moves further away, however, it is difficult to qualitatively describe the local structure except to say that it gradually transforms until one reaches another high symmetry site, e.g. site 9, which is approximately tetrahedral. Here, the levels split into two a singlet with two doublets above and below it. While this is largely consistent with row 9 of table 2, one can see some discrepancies. The perturbative model suggests that Fe should have zero magnetization, but DFT calculation instead yields 2 $\mu_B$. Additionally, Ni has zero magnetization in DFT compared to the model prediction of 2 $\mu_B$, although no magnetization was found for Ni in any configuration. These disagreements can be understood as this simple model neglects 4s states, which in the DFT calculation may be promoted to the d-orbitals which then become fully occupied. Despite this, we find that the site-dependent spin magnetic moment of the TM intercalated impurities can largely be explained due to differences in local symmetry leading to different d-orbital splitting.

\section{Conclusion}

The moir\'e pattern formed by two hexagonal lattices can be analyzed using the Moir\'e Condition Equation. The result is that essentially all symmetrically distinct moir\'e patterns take the form $(n,m,m,n)$, in which $(0,0)$ is fixed and the atom of one layer at position $(n,m)$ is rotated to coincide with the position $(m,n)$ of the other layer. The number of moirons that appear within each supercell is determined by the difference $n-m$, and is given by $N=(n-m)^2$. Furthermore, the real-space method can be extended to deal with lattices that do not have the same lattice constant.

Concerning the systems with a transition metal in the intercalated space of a twisted h-BN twisted bilayer, we conclude that zinc does not form a stable configuration, while all other transition metals at all other sites are stable but at a higher energy per transition metal than bulk. Therefore, at a very high density of these atoms we would expect the atoms to clump. For transition metals such a vanadium, which shows a large energy difference between being in a aligned/anti-aligned state vs an unaligned state (0.94 eV difference), these atoms would tend to form a pattern corresponding with the moir\'e pattern as their density is increased. Also, most of the transition metals have various magnetic state, except for nickel (and zinc), which are a function of their location within the Moir\'e lattice. The dependence of the magnetization state on local environment is mostly described by a crystal field splitting model.


\end{multicols}
\FloatBarrier
\bibliography{hBN_Moire_transitonmetal}
\bibliographystyle{ieeetr}

\end{document}